\documentclass[12pt,a4paper, twoside]{article}
\usepackage{epsf,multicol,ifthen}
\usepackage[english]{babel}
\usepackage[dvips]{graphicx}
\usepackage{cite}
\usepackage{hyperref}
\newcommand{\bec}{\begin{center}}
\newcommand{\ec}{\end{center}}
\newcommand{\bee}{\begin{equation}}
\newcommand{\ee}{\end{equation}}

\title{CENTRAL CHARGE AND TOPOLOGICAL INVARIANT OF CALABI-YAU
MANIFOLDS }

\author{T.V. Obikhod\thanks{E-mail: obikhod@kinr.kiev.ua}\\
\small\emph{Institute for Nuclear Research, National Academy of Science of Ukraine} \\ 
\small\emph{47, prosp. Nauki, Kiev, 03028, Ukraine}}

\begin{document}

\maketitle
UDK 514.83
\abstract{F-theory, as a 12-dimensional theory that is a contender of the Theory of Everything, should
be compactified into elliptically fibered threefolds or fourfolds of Calabi-Yau.
Such manifolds have an elliptic curve as a fiber, and their bases
may have singularities. 
We considered orbifold as simplest non-flat construction. 
Blow up modes of orbifold singularities can be considered as 
coordinates of complexified Kahler moduli space. Quiver diagrams are used for discribing 
D-branes near orbifold point. 
In this case it is possible to calculate Euler character defined through
 $\mbox{Ext}^i(A,B)$ groups and
coherent sheaves $A, B$ over projective space,
which are representations of orbifold space after blowing up procedure.
These fractional sheaves are characterized by D0, D2 and D4 Ramon-Ramon charges,
which have special type, calculated for $C^3/Z_3$ case.
BPS central charge for $C^3/Z_3$ orbifold is calculated through
Ramon-Ramon charges and Picard-Fuchs periods.}\\
\vspace{5mm}

\newpage

\section{Introduction}
Modern high energy theoretical physics is a unified theory of
all particles and all interactions. It is Theory of Everything, 
because it gives a universal description of the
processes occurring on modern accelerators, and processes in the Universe.

  Theory of everything (abbr. TOE) - hypothetical
combined physical and mathematical theory describing all known
fundamental interactions. This theory unifies all four
fundamental interactions in nature. The main problem of building
TOE is that quantum mechanics and general
theory of relativity have different applications.
Quantum mechanics is mainly used to describe the microworld, and
general relativity is applicable to the macro world. 
But it does not mean that such theory cannot be constructed.

  Modern physics requires from TOE the unification of four
 fundamental interactions: \\
$\bullet $ gravitational interaction; \\
$\bullet $ electromagnetic interaction; \\
$\bullet $ strong nuclear interaction; \\
$\bullet $ weak nuclear interaction.

  The first step towards this was the unification of the electromagnetic and
weak interactions in the theory of electro-weak interaction created by
in 1967 by Stephen Weinberg, Sheldon Glashow and Abdus Salam.
In 1973, the theory of strong interaction was proposed.

The main candidate as TOE
is F-theory, which operates with a large number of dimensions.
Thanks to the ideas of Kaluza and Klein it became
 possible to create theories operating with large extra dimensions.
The use of extra dimensions prompted the answer to the question about
why the effect of gravity appears much weaker than
other types of interactions. The generally accepted answer is that
gravity exists in extra dimensions, therefore its effect on
observable measurements weakened.

  F-theory is a string twelve-dimensional theory defined on
energy scale of about 10$^{19}$~GeV \cite {1.}. F-theory 
compactification
leads to a new type of vacuum, so to study supersymmetry we
must compactify the F-theory on Calabi-Yau manifolds.
Since there are many Calabi-Yau manifolds, we are dealing with
a large number of new models implemented in low-energy
approximation. Studying the singularities of Calabi manifold
determines the physical characteristics of topological solitonic
states which plays the role of particles in high energy physics.

    Compactification of F-theory on different 
Calabi-Yau manifolds allows to calculate topological invariants.

  Let us consider in more detail the compactification of F-theory on threefolds Calabi Yau.

\section{Calabi-Yau threefold compactification}
Twelve-dimensional space describing space-time
and internal degrees of freedom, we compactify as follows:
\[R^6 \times X^6 \ ,\]
where $R^6$ - six-dimensional space-time, on which acts
conformal group SO(4, 2), and $X^6 $ - threefold, 
which is three-dimensional
Calabi Yau complex manifold \cite{2.}. 
\subsection{Toric representation of threefolds}

Let's consider weighted
projective space defined as follows:
\[P^4_{\omega_1,\ldots,\omega_5 }=P^4/Z_{\omega_1}\times \ldots \times Z_{\omega_5}\ , \]
where $P^4$ - four-dimensional projective space,
$Z_{\omega_i}$ - cyclic group of order $\omega_i$. 
On weighted projective space $P^4_{\omega_1,\ldots,\omega_5 }$
is defined polynomial $W(\varphi_1, \ldots, \varphi_5)$,
called superpotential which satisfies the homogeneity condition
\[W(x^{\omega_1}\varphi_1, \ldots, x^{\omega_5}\varphi_5)=x^d W(\varphi_1, \ldots, \varphi_5)\ ,\]
where $d=\sum\limits_{i=1}^5\omega_i$, $\varphi_1, \ldots, \varphi_5 \in P^4_{\omega_1,\ldots,\omega_5 } $.
The set of points $p\in P^4_{\omega_1,\ldots,\omega_5 } $,
satisfying the condition $W(p)=0$ forms Calabi-Yau threefold
$X_d(\omega_1, \ldots, \omega_5)$ .

	The simplest examples of toric varieties \cite{3.} are projective spaces.
Let's consider $P^{2}$ defined as follows:
\[P^{2} = \frac{C^{3}/{0}}{C/{0}}, \]
where dividing by $ C/{0}$ means identification of points connected
by equivalence relation
\[(x, y, z)\sim(\lambda x, \lambda y, \lambda z) \]
\[\lambda \in C/{0}, \]
$ x, y, z $ are homogeneous coordinates. Elliptic curve in $P^{2}$
is described by the Weierstrass equation
\[y^2z = x^3 + axz^2 + bz^3. \] 
In general Calabi-Yau  manifold can be described by Weierstrass form
\[y^2=x^3+xf+g,\]
which describes an elliptic fibration (parametrized
by $(y, x)$) over the base, where $f, g$ - functions defined on the base. 
In some divisors $D_i$ the layer are degenerated. Such divisors are zeros 
of discriminant 
\[\Delta=4f^3+27g^2.\]
The singularities of Calabi-Yau manifold are singularities of its
elliptic fibrations. These singularities are
coded in polynomials $f, g$ and their type determines the gauge group and
matter content of compactified F-theory.

  The classification of singularities of elliptic fibrations was given by Kodaira and presented table 1.\\
  \bec
\emph{\textbf{Table 1.}} {\it Kodaira classification of singularities of elliptic fibrations}
\begin{tabular}{|c|c|c|}\hline
$ord(\Delta)$&Type of fiber & Type of singularity \\ \hline
0&smooth&no\\ \hline
n&$I_n$&$A_{n-1}$ \\ \hline
2&$II$&no \\ \hline
3&$III$&$A_{1}$ \\ \hline
4&$IV$&$A_{2}$ \\ \hline
n+6&$I^*_n$&$D_{n+4}$ \\ \hline
8&$IV^*$&$E_{6}$ \\ \hline
9&$III^*$&$E_{7}$ \\ \hline
10&$II^*$&$E_{8}$ \\ \hline
\end{tabular}
\ec
\vspace*{3mm}
The classification of elliptic fibers is presented in Figure 1.
\begin{figure}[ht]
\centerline{\includegraphics[width=0.38\textwidth]{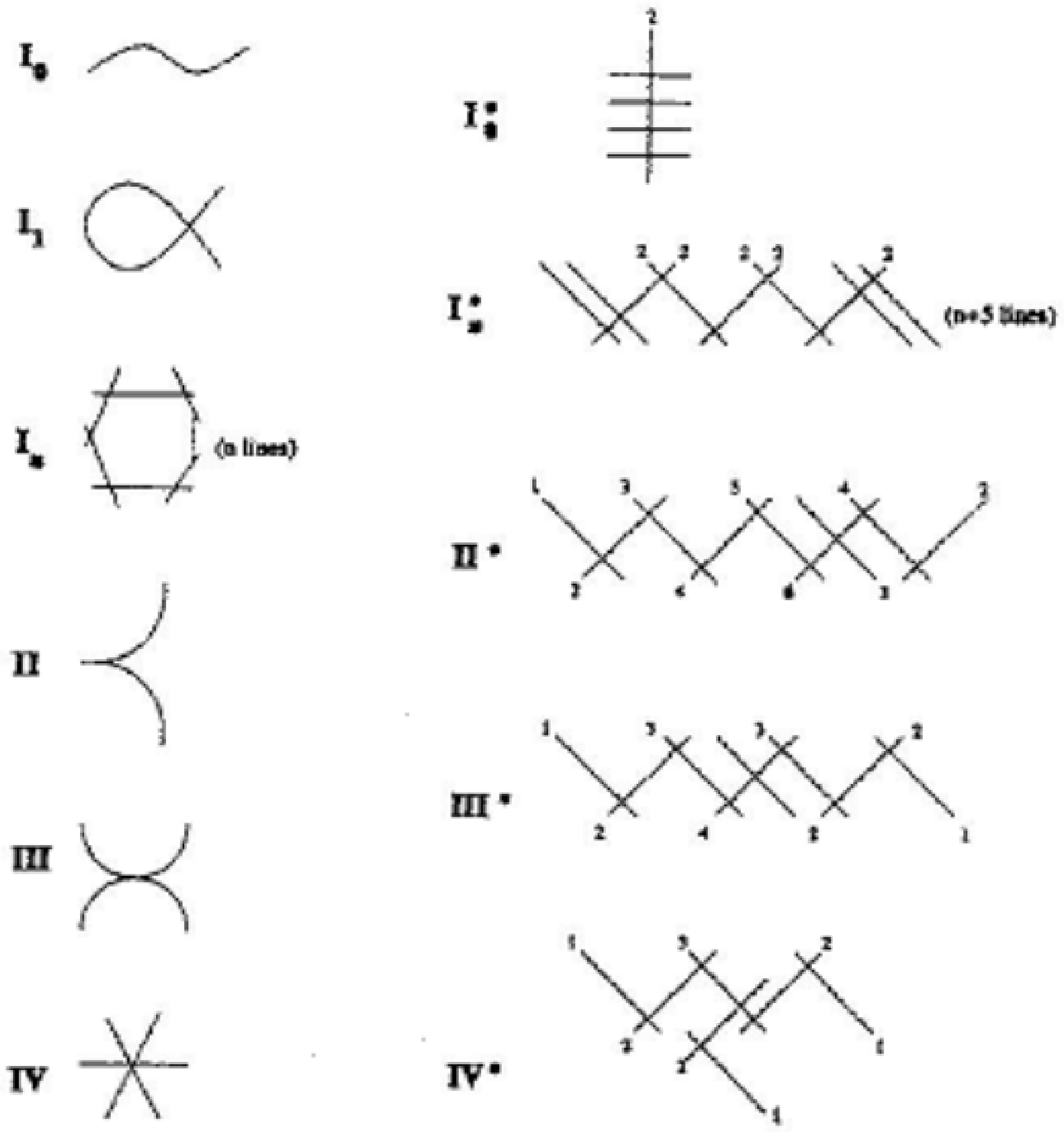}}
\bec\bf{Fig.1. The classification of elliptic fibers.}\ec
\end{figure}

\section{Calculation of topological invariants}
\subsection{Ramon-Ramon charges}
One of the most interesting problems of modern high-energy physics is the calculation of topological invariants - analogs of high-energy observables in physics. In this aspect, symmetries and the use of the apparatus of algebraic geometry play an indispensable role. We considered orbifold as simplest non-flat constructions. For D3-branes on such internal space 
$C^n/\Gamma$ the representations are characterized by gauge groups 
$G=\oplus_iU(N_i)$. In this case the superpotential is of N=4 $U(N)$ super
Yang-Mills, 
\[W_{N=4}=\mbox{tr}X^1[X^2,X^3],\]
where $X^i$ are chiral matter fields in production of fundamental
representation $V^i\cong C^{N_i}$ of the group $U(N_i)$. 
Blow up modes of orbifold singularities can be considered as coordinates of complexified Kahler moduli space. Quiver diagrams are used for discribing 
D-branes near orbifold point. 

\begin{figure}[ht]
\centerline{\includegraphics[width=0.38\textwidth]{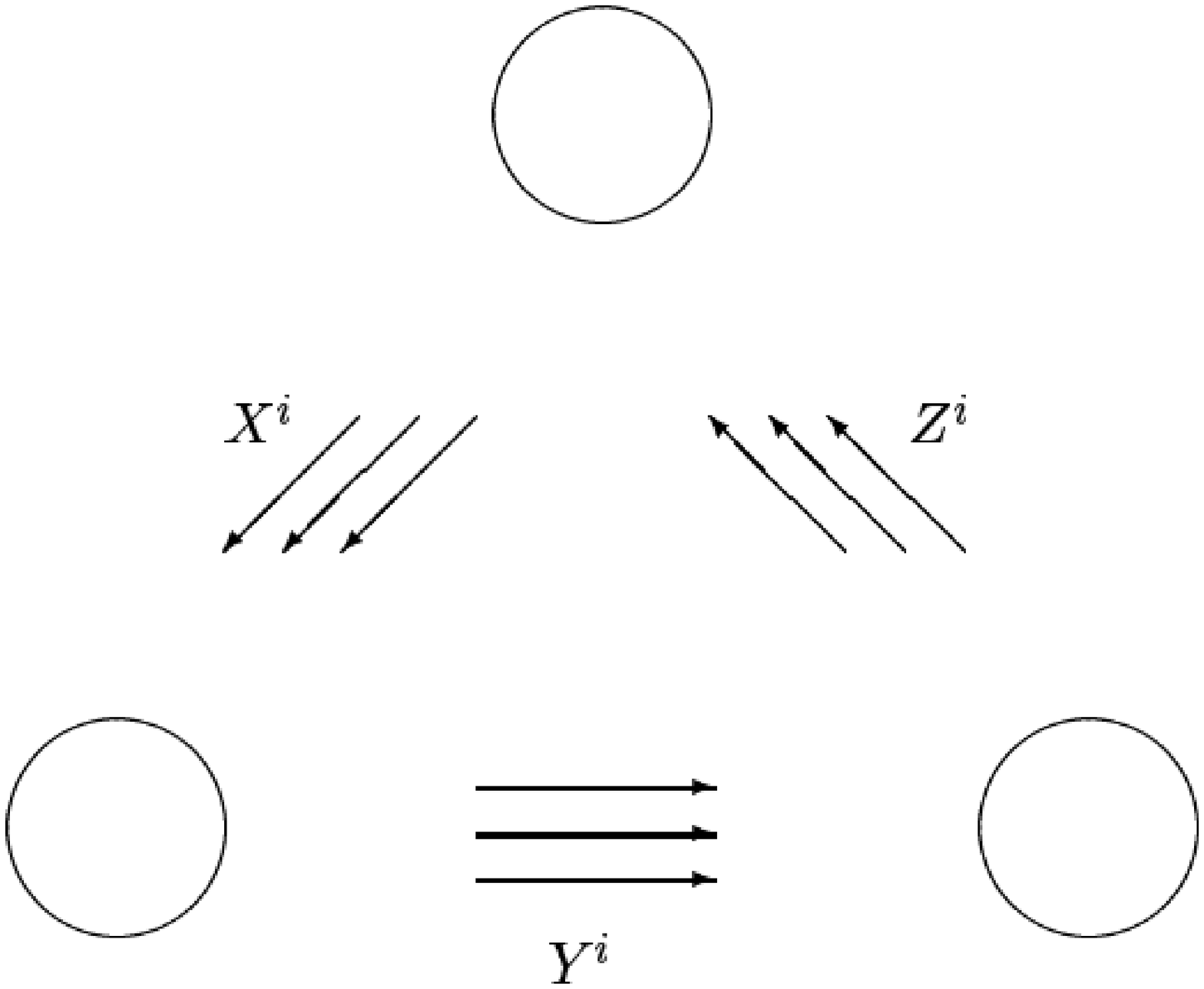}}
\bec\bf{Fig.2. The $C^3/Z_3$ quiver.}\ec
\end{figure}

In this case it is possible to calculate Euler character defined as
\[\chi(A,B)=\sum_i(-1)^i\mbox{dimExt}^i(A,B),\]
where $\mbox{Ext}^0(A,B)\equiv \mbox{Hom}(A,B)$ and
$A, B$ are coherent sheaves over projective space, $P^N$ (general case),
which are representations of orbifold space after blowing up procedure.

Since we will deal with orbifolds $C^3/Z_3$ in the future, it is necessary to emphasize the following equivalence relation
\[(x_1x_2x_3)\sim(e^{2i\pi/3}x_1, e^{2i\pi/3}x_2, e^{2i\pi/3}x_3), \ e^{2i\pi/3}\in Z_3\]
Orbifold is not a manifold, since it has singularities at a point $(0, 0, 0)$. Blowing up the singularity of the orbifold $C^3/Z_3$, we obtain a sheave
${\cal{O}}_{P^2}(-3)$ with which we will work further.
In particular, the Euler matrix for sheaves ${\cal{O}}_{P^2}$, 
${\cal{O}}_{P^2}(1)$, ${\cal{O}}_{P^2}(2)$
over projective space, $P^2$ looks like 
\[ \chi({\cal{O}}_{P^2}(1), {\cal{O}}_{P^2}(2)) = \left( \begin{array}{ccc}
1 & 3& 6 \\
0& 1 &3 \\
0 & 0 & 1 \end{array} \right).\] 
\vspace{2mm}
Transposed matrix has the form
\bigskip 
\[ \left( \begin{array}{ccc}
1 & 3& 6 \\
0& 1 &3 \\
0 & 0 & 1 \end{array} \right)
\Rightarrow
\left( \begin{array}{ccc}
1 & 0& 0 \\
3& 1 &0 \\
6& 3 & 1 \end{array} \right).\] 
\vspace{2mm}
The rows of matrices are RR-charges characterizing the sheaves:
\begin{equation}
{\cal{O}}_{P^2}(-3)=(6\ 3\ 1), {\cal{O}}_{P^2}(-2)=(3\ 1\ 0), 
{\cal{O}}_{P^2}(-1)= (1\ 0\ 0), 
\end{equation}
\begin{equation}
{\cal{O}}_{P^2}=(0\ 0\ 1), {\cal{O}}_{P^2}(1)=(0\ 1\ 3), 
{\cal{O}}_{P^2}(2)= (1\ 3\ 6),
\end{equation}
\vspace{2mm}
which can be written through large volume charges $(Q_4, Q_2, Q_0)$:
\[Q_4=n_1-2n_2+n_3, \ \ Q_2=-n_1+n_2, \ \ Q_0=\frac{n_1+n_2}{2}\]
included in the definition of the  Chern character $ch(n_1n_2n_3)$
\[ch(n_1n_2n_3)=Q_4+Q_2w+Q_0w^2,\]
where $w$ - Wu number.
Then sheaves (1), (2) describe fractional branes \cite{4.}
\[{\cal{O}}_{P^2}(-3)=(1\ -3\ \ \frac{9}{2}), {\cal{O}}_{P^2}(-2)=(1\ -2\ \ \frac{4}{2}), {\cal{O}}_{P^2}(-1)= (1\ -1\ \ \frac{1}{2}), \]
\[{\cal{O}}_{P^2}=(1\ 0\ 0), {\cal{O}}_{P^2}(1)=(1\ 1\ \ \frac{1}{2}), 
{\cal{O}}_{P^2}(2)= (1\ 2\ \ \frac{4}{2}),\]
General formula for Chern character of bundle $E$:
\[ch(E)=k+c_1(E)+\frac{1}{2}(c_1(E)^2-2c_2(E))+\ldots ,\]
where $c_i(E)$ are the Chern classes of line bundle $E$.
In our case of a line bundle ${\cal{O}}_{P^2}(k)$, only the first Chern class is nonzero, and therefore the formula for the Chern character is following
\begin{equation}
ch(E)=k+c_1(E)+\frac{1}{2}c_1^2
\end{equation}
As 
\[1+c_1(E)+\ldots + c_n(E)=\prod\limits_{i=1}^{n}(1+w_i),\]
then $c_1(E)=w_1=w$ and formula (3) can be rewritten
\begin{equation}
ch(n_1n_2n_3)=Q_4+Q_2w+Q_0w^2,
\end{equation}
where Ramon-Ramon charges $(n_1n_2n_3)$ 
characterize the bundle $E$, the rank of the line bundle $Q_4=1, Q_2=c_1$ by the fundamental cycle, $Q_0=\frac{c_1^2}{2}$ from a comparison of formulas (3) and (4). 

	Thus fractional sheaves ${\cal{O}}_{P^2}(k)$ are characterized by $Q_0, Q_2, Q_4$ Ramon-Ramon charges, which have special type, calculated for $C^3/Z_3$ case.

\subsection{BPS central charge}
As we are interested in the moduli spaces, we give them a visual definition.
Suppose we have a cube curve with the parameter $\lambda$
\begin{equation}
y^2-x(x-1)(x-\lambda)=0
\end{equation}
As $\lambda$ - the variable value, then the equation (5) describes a continuous family of cubic curves. The parameter spaces describing continuous families of manifolds are called moduli spaces. We form $\frac{dx}{y}$ the form where $y$ are determined from equation (5). It turns out that periods
$\pi_1(\lambda),\ pi_2(\lambda)$:
\[\pi_1(\lambda)=2\int\limits_0^1\frac{dx}{[x(x-1)(x-\lambda)]^{1/2}}, \ \
\pi_2(\lambda)=2\int\limits_1^{\lambda}\frac{dx}{[x(x-1)(x-\lambda)]^{1/2}}\]
satisfy Picard-Fuchs equation
\begin{equation}
\frac{1}{4}\pi_i+(2\lambda -1)\frac{d\pi_i}{d\lambda}+\lambda(\lambda -1)
\frac{d^2\pi_i}{d\lambda^2}=0\ .
\end{equation}
Periods that satisfy equation (6) describe the moduli space of a cubic curves.
For the moduli space of a line bundle ${\cal{O}}_{P^2}(-3)$, Picard-Fuchs equation and its solutions are written as
\[
\Biggl(z\frac{d}{dz}\Biggr)^3+27z\Biggl(z\frac{d}{dz}\Biggr)\Biggl(z\frac{d}{dz}+\frac{1}{3}\Biggr)\Biggl(z\frac{d}{dz}+\frac{2}{3}\Biggr)\Pi=0
\]
\[
\Pi_0=1,
\]
\[\Pi_1=\frac{1}{2i\pi}log\ z=t=w_0,\]
\[\Pi_2=t^2-t-\frac{1}{6}=-\frac{2}{3}(w_0-w_1) \ .\]
The BPS central charge \cite{5.} associated with the D-brane over $C^3/Z_3$ with Ramon-Ramon-charge $n=(n_1n_2n_3)$ and with the Picard-Fuchs period 
$\Pi=(\Pi_0\Pi_1\Pi_2)$ is given by the formula
\[Z(n)=n\cdot\Pi\]
The central charge associated with the sheave ${\cal{O}}_{P^2}(k)$ is given by the formula 
\[Z({\cal{O}}_{P^2}(k))=-(k+\frac{1}{3}w_0)+\frac{1}{3}w_1+\frac{1}{2}k^2 +
\frac{1}{2}k+\frac{1}{3}\ .
\]
\section{Conclusion}
In the framework of F-theory we prsented the ideology of extra dimensional spaces. It was stressed the exceptional role of topological invariants for Calabi-Yau manifolds. We have considered the special type of the space of extra dimensions - orbifold $C^/Z_3$. Using blowing up procedure of singularity we calculated special type of topological invariant - Ramon-Ramon central charges of fractional sheaves, in which is encoded the information about the structure of line bundles. Consideration of moduli space of orbifold leads us to the equation of Picard-Fuchs periods, through which we calculated central charge for sheave ${\cal{O}}_{P^2}(k)$. This topological invariant is of importance because of information of stability of D-branes as bound states of fractional branes or sheaves presented in this paper.


\begin{thebibliography}{99}

\bibitem{1.}C. Vafa, {\it Evidence for F-theory}, arXiv: hep-th/9602022; \\
D. R. Morrison and C. Vafa, {\it Compactifications of F-theory on Calabi-Yau threefolds (I)}, 
Nucl. Phys. B473 (1996) 74; \\
D. R. Morrison and C. Vafa, {\it Compactifications of F-theory on Calabi-Yau threefolds (II)}, 
Nucl. Phys. B476 (1996) 437.

\bibitem{2.} S. Hosono , A. Klemm , S. Theisen , S.-T. Yau, {\it Mirror symmetry, mirror map and applications
to complete intersection Calabi-Yau spaces}, Nucl. Phys. B433 (1995) 501.

\bibitem{3.} V. V. Batyrev, {\it Variations of the Mixed Hodge Structure of Affine Hypersurfaces in Algebraic
Tori}, Duke Math. J. 69, (1993), 349-409.

\bibitem{4.}
  D. Diaconescu and J. Gomis
   \emph{Fractional branes and boundary states in orbifold theories},
  JHEP 0010 (2000) 001, hep-th/9906242.
\bibitem{5.}
  A. Klemm , P. Mayr, C. Vafa, {\it BPS states of exceptional non-critical strings}, Harvard, 1996. – 29 p. – (Preprint, HUTP-96/A031).

\end{thebibliography}
\end{document}